\documentclass[prl,aps,twocolumn,showpacs,amssymb]{revtex4}

\usepackage{epsfig}
\usepackage{times}

\newcommand{\neswarrow}{\text{$\nearrow$\llap{$\swarrow$}}}
\newcommand{\nwsearrow}{\text{$\nwarrow$\llap{$\searrow$}}}

\begin{document}

\title{Experimental demonstration of
entanglement-enhanced classical
communication over~a~quantum~channel with correlated noise}

\author{Konrad Banaszek}
\affiliation{Clarendon Laboratory, University of Oxford,
Parks Road, Oxford OX1 3PU, United Kingdom}

\author{Andrzej Dragan, Wojciech Wasilewski, and Czes{\l}aw Radzewicz}
\affiliation{Wydzia{\l} Fizyki, Uniwersytet Warszawski, ul.\ Ho\.{z}a 69,
PL-00-681 Warszawa, Poland}

\begin{abstract}
We present an experiment demonstrating entanglement-enhanced classical communication capacity of a quantum channel with correlated noise. The channel is modelled by a fiber optic link exhibiting random birefringence that fluctuates on a time scale much longer than the temporal separation between consecutive uses of the channel. In this setting, introducing entanglement between two photons travelling down the fiber allows one to encode reliably up to one bit of information into their joint polarization degree of freedom. When no quantum correlations between two separate uses of the channel are allowed, this capacity is reduced by a factor of more than three. We demonstrated this effect using a fiber-coupled source of entagled photon pairs based on spontaneous parametric down-conversion, and a linear-optics Bell state measurement.
\end{abstract}


\maketitle

Quantum mechanics admits existence of correlations that cannot be explained simply as statistical uncertainty in assigning definite realistic properties locally to each of the subsystems. This feature of quantum mechanics was quantified first in the form of Bell's inequalities that discriminate it against theories based on the assumption of local realism \cite{BellPhys65}. In recent years, this specifically quantum form of correlations, known commonly as entanglement, is being exploited to develop novel modes of information processing that offer advantages not available through the classical approach \cite{NielsenChuang}.
The well-known examples include solving certain computation tasks via collective unitary operations on coherent registers of quantum particles \cite{QComputing}, or distributing a cryptographic key with the security verified by the detection of entanglement \cite{QCrypto}. Entanglement plays also a non-trivial role in a number of scenarios for sending both classical information and quantum states over noisy quantum channels \cite{BennShorIEEE98}.

In this Letter, we demonstrate experimentally how entanglement can be used to
enhance classical communication over a noisy channel. Our experiment follows  recent theoretical studies \cite{MaccPalmPRA02,BoweMancPRA04,BallDragPRA04}
that analyzed classical communication over quantum channels in which the noise affecting consecutive uses is correlated. It is then natural to consider two types of input ensembles used for communication. The first one is restricted only to introducing classical correlations between separate uses of the channel, and it can be prepared by adjusting individually the quantum state of each particle sent through the channel in a single use. The second, completely general input ensemble includes entangled states of many particles that are subsequently sent through the channel one-by-one. Theoretical analysis of model situations showed that indeed the use of entangled input ensemble can substantially enhance the classical capacity, and our work provides a proof-of-principle experimental verification of this prediction.

Our experiment implements the idea described in Ref.~\cite{BallDragPRA04} which we will now briefly review. The model of a noisy quantum channel analyzed there
was motivated by fluctuating birefringence of a standard optical fiber, which scrambles the polarization of an input light pulse to a completely mixed state. However, 
the characteristic time scale of birefringence fluctuations is usually much longer than the temporal separation between consecutive light pulses. The fact that neighboring light pulses undergo a random but nearly identical polarization transformation opens up the possibility of encoding information in the polarization degree of freedom of the transmitted light. Let us consider a simple scenario, when each of the two pulses contains exactly one photon, which is the minimum amount of light required
to trigger a definite reponse on an ideal, noise-free photon counting detector. If the sender does not have technical capabilities to generate entangled photon pairs, all she can do is to assign a definite polarization to each of the photons. In this setting, the optimal encoding of information turns out to be a pair of photons with polarizations that are either identical or mutually orthogonal, for example 
$|\neswarrow\neswarrow\rangle$ and $|\neswarrow\nwsearrow\rangle$. (The diagonal polarization basis introduced here will be natural in the description of our experiment.) However, after collective polarization scrambling these two states cannot be discriminated unambiguously, and the highest achievable transfer rate is $\log_2(5/4)\approx 0.322$ bits per two photons \cite{BallDragPRA04}. 

The advantage of employing entangled states in the above scenario becomes obvious when we recall that the singlet polarization state of two photons 
$|\Psi_-\rangle = (|\neswarrow\nwsearrow\rangle - |\nwsearrow\neswarrow\rangle)/\sqrt{2}$
remains invariant under correlated depolarization
\cite{KwiaBergSCI00}, and therefore it can be discriminated unambiguously against the orthogonal triplet subspace. This fact allows one to encode faithfully one full bit of information into the polarization state of two photons by sending either the singlet state or one of the triplet states, which represents more than a three-fold improvement over the separable case. 
A detailed calculation based on the Holevo bound shows that this is the optimal capacity attainable with two-photon entangled states \cite{BallDragPRA04}.
The measurement required on the receiver side  to maximize the capacity in both the entangled and the separable cases has the form of a two-outcome projection either onto the one-dimensional antisymmetric subspace spanned by the singlet state $|\Psi_-\rangle\langle\Psi_-|$, or onto the three-dimensional symmetric triplet subspace $\hat{\openone}-|\Psi_-\rangle\langle\Psi_-|$.  

We demonstrated the effect of enhanced communication capacity using a parametric down-conversion source of polarization-entangled photon pairs introduced by Kwiat {\em et al.} \cite{KwiaWaksPRA99} which consists
of two type-I nonlinear crystals with optical axes lying in perpendicular planes. The noisy fiber optic link was modelled by a single-mode fiber suspended between several mechanical pendulums that were excited at regular intervals during the measurements. Finally, the measurement of the output state was performed with the Bell-state analyzer based on linear optics suggested by Braunstein and Mann \cite{BrauMannPRA95}.

\begin{figure}
\epsfig{file=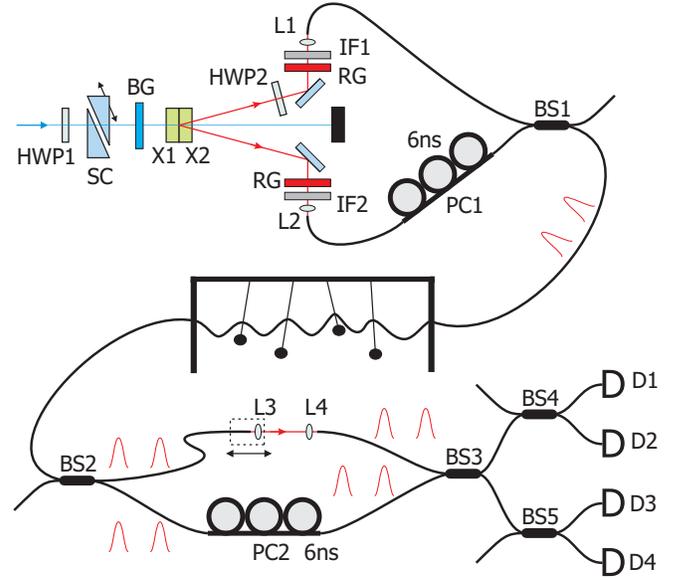,width=3.375in}
\caption{Experimental setup. HWP1, HWP2, half-wave plates; SC, Soleil compensator; BG, blue BG39 filter; X1, X2, down-conversion crystals;
RG, red RG665 filters;
IF1, IF2, interference filters; L1, L2, aspheric lenses; PC1, PC2, manual polarization controllers introducing additional 6~ns delay. BS1, \ldots , BS5 single-mode 50/50 fiber optic couplers; L3, L4; collimators; D1, \ldots , D4, photon counting modules. Singlet events are coincidences between D1\&D3, D1\&D4, D2\&D3, or D2\&D4, whereas triplet events are defined as concidences between D1\&D2 or D3\&D4.}
\label{Fig:Setup} \end{figure}

Details of the experimental setup are depicted schematically in Fig.~\ref{Fig:Setup}. We start by doubling the output of a mode-locked Ti:sapphire oscillator to obtain a train 
of ultraviolet pulses with 390~nm central wavelength, 3.5~nm FWHM bandwidth, 16~mW average power, and 78~MHz repetition rate.
Before arriving at the down-conversion crystals, the pump pulses pass through a half-wave plate HWP1, a Soleil compensator and a BG39 blue color filter removing residual fundamental light. The half-wave plate HWP1 distributes the pump beam between two linear and orthogonal polarization components pumping each of the down-conversion crystals. The purpose of the Soleil compensator, made of two indentically cut quartz wedges properly oriented with respect to the axes of the down-conversion crystals, is two-fold. On the coarse scale, it is used to compensate for the temporal delay between the photons generated in the first and the second crystals \cite{NambUsamPRA02}. On the fine scale, it provides means to adjust the relative phase between the two components in the generated maximally entangled polarization state.

After the preparation stage the UV beam, focused to the diameter of 250~$\mu$m (FWHM), pumps a pair of type-I 1-mm thick $\beta$-barium borate crystals. Both crystals are identically cut at an angle that corresponds to the frequency-degenerate down conversion process taking place on a cone with the half opening angle of 1.38$^\circ$. The down-converted photon pairs are collected in the horizontal plane, parallel to the optical table surface, while the mutually perpendicular planes containing the optical axes of the two crystals are rotated by $45^{\circ}$ with respect to the optical table. This specific geometry of the experiment was motivated by a theoretical study that indicated such an arrangement as giving optimal indistinguishability of photon pairs produced in both the crystals by a focused, broadband pump beam \cite{DraganInPreparation}.
In this arrangement the components to the two-photon wave function contributed by the first and the second crystal are given respectively by $|\neswarrow\neswarrow\rangle$ and $|\nwsearrow\nwsearrow\rangle$ in the reference frame of the optical table.
The down-converted photons, after passing through 10.5~nm bandwidth interference filters centered at 780~nm and RG665 long pass filters, are coupled by aspheric lenses into single-mode fibers. 

As the input ensembles for the optimal communication protocol through the depolarizing fiber we select the pairs: $|\neswarrow\neswarrow\rangle$ and $|\neswarrow\nwsearrow\rangle$ in the separable case,
and $(|\neswarrow\neswarrow\rangle + |\nwsearrow\nwsearrow\rangle)/\sqrt{2}$
and $(|\neswarrow\nwsearrow\rangle - |\nwsearrow\neswarrow\rangle)/\sqrt{2}$ in the entangled case. The transformation between the states within each pair is performed using the half-wave plate HWP2 placed in the path of one of the down-converted photons. After coupling into the fibers, we introduce a relative 6~ns temporal delay between the photons with the help of an additional fiber wound on the polarization controller PC1. Then the photon pairs are combined into a single spatial mode by the 50/50 coupler BS1 which launches them into the depolarizing fiber modelling a noisy quantum channel. This completes the preparation procedure on the transmitter side.

On the output of the depolarizing fiber the photons are separated with the help of the 50/50 coupler BS2. The measurement will select the cases when the photon arriving earlier is delayed by 6~ns in an additional fiber cable in order to restore the temporal overlap within a pair. A computer-controlled optical delay line built on a motorized translation stage (Physik Instrumente M-014.D01) provides fine tuning of the delay. With matched arrival times, the polarization state of the photon pair can be determined by sending it to the balanced coupler BS3. The photons emerge on two separate output ports of the coupler BS3 only if the pair was initially in the singlet state $|\Psi_-\rangle$ \cite{BrauMannPRA95}. Detecting both the photons in the same output port of BS3 corresponds to the projection on the symmetric triplet subspace $\hat{\openone}-|\Psi_-\rangle\langle\Psi_-|$. As the detectors used in our setup were standard single-photon counting modules (PerkinElmer SPCM-AQR-14-FC) based on avalanche photodiodes operated in the Geiger mode that cannot resolve multiphoton detection events, we followed the coupler BS3 by two more couplers BS4 and BS5 terminated with photon counting modules $\text{D1},\ldots ,\text{D4}$. Excluding detector losses, this gives a 50\%  chance of detecting the
$\hat{\openone}-|\Psi_-\rangle\langle\Psi_-|$ projection
as a two-fold coincidence between either D1 and D2 or D3 and D4. We will label these events as triplet coincidences. Projection onto the singlet state $|\Psi_-\rangle\langle\Psi_-|$ is heralded by a coincidence between any of four remaining pairs of detectors. Electronic signals from the photon counting modules are processed using standard NIM electronics, and two-fold coincidences between all pairs of detectors are counted using a PCI card (National Instruments PCI-6602). Polarizations controllers PC1 and PC2 ensure that the photons experience the same polarization transformation on their spatially distinct paths from the aspheric lenses to the coupler BS1, and between the couplers BS2 and BS3. 

Launching the photon pairs into the depolarizing channel and their separation on the output is realized in our setup with passive fiber optic couplers rather than active fast optical switches which are not readily available in the 780~nm wavelength region. This has two consequences. First, each of the couplers directs photon pairs into required output ports only in 25\% of cases, assuming no excess losses. Second, detectors register single photons that have passed through paths in the setup other than those leading to the temporal overlap at the coupler BS3. This gives single counts occuring 6~ns before as well as after the temporal slot of interest, which however lie well beyond the coincidence window set to 3~ns. The only possibility that these events could yield a coincidence is the generation of two photon pairs by the consecutive pump pulses, which is highly unlikely in our setup due to low pump power. 

\begin{figure}
\epsfig{file=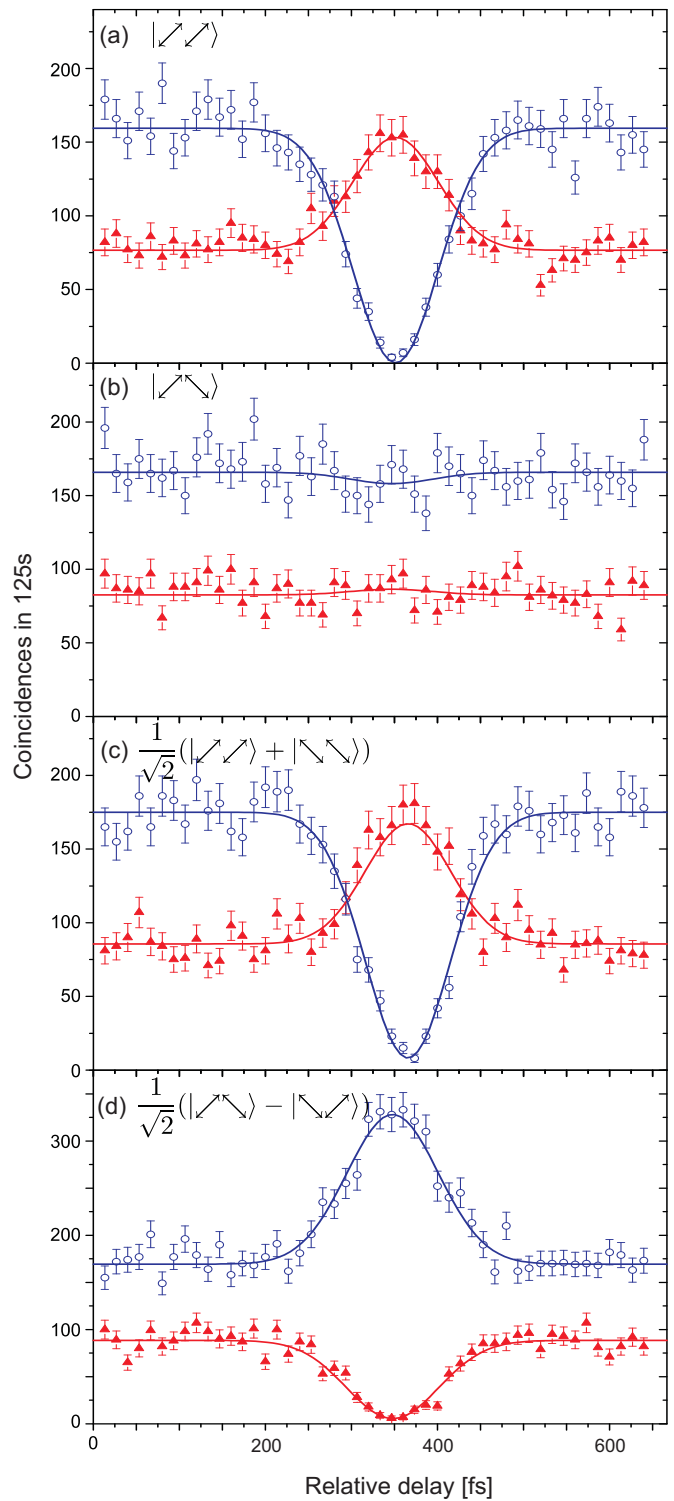}
\caption{Singlet ($\circ$) and triplet ($\blacktriangle$)
coincidences as a function of the optical delay for separable states of (a) parallel and (b) orthogonal polarizations, and for entangled triplet (c) and singlet (d) polarization state. The solid lines are fitted Gaussian functions.}
\label{Fig:GoryiDoliny}
\end{figure}

We measured the number of coincidences of both types --- singlet and triplet --- as a function of the displacement of the translation stage. Experimental results are shown in Fig.~\ref{Fig:GoryiDoliny}.
The first series of measurements was taken with the half-wave plate HWP1 oriented to produce photon pairs only in one of the down-conversion crystals, and the axis of the half-wave plate HWP2 oriented in the diagonal basis. The polarizations
of the down-converted photons are then identical, the state launched into the fiber is
$|\neswarrow\neswarrow\rangle$, and we observe a suppression of singlet coincidences for matched arrival times, as seen in Fig.~\ref{Fig:GoryiDoliny}(a).
This is the standard effect of bunching in two-photon Hong-Ou-Mandel interference
\cite{HongOuPRL87}. The second of the states in the optimal separable ensemble is obtained by rotating the half-wave plate HWP2 by $45^\circ$ and thus generating $|\neswarrow\nwsearrow\rangle$. Photons in these pairs remain fully distinguishable throughout the setup and they do not interfere at BS3. Consequently, the probabilities of obtaining a singlet or a triplet coincidence are equal, assuming ideal photon number resolution.

In the second series of measurements, we used entangled singlet and triplet states. They were generated and characterized as follows. First, half-wave plates and polarizing beam splitters were placed in front of the interference filters, and the ends of the collecting fibers were connected directly to a pair of photon counting modules. Then
the polarization of the pump beam was rotated by the half-wave plate HWP1 to balance the number of coincidences generated by the first and the second crystal, and the Soleil compensator was aligned to minimize the number coincidences when only pairs of horizontally polarized photons were detected.
The polarization state of the photon pairs leaving the crystal was then $(|\neswarrow\neswarrow\rangle - |\nwsearrow\nwsearrow\rangle)/\sqrt{2}$.
We also tested the indistinguishability of photon pairs produced in the crystals. This was done by measuring the visibility of the polarization correlations in the horizontal-vertical basis and yielded  a figure exceeding 98\%.
Both polarizing beam splitters and one half-wave plate were removed before the photon pair was launched into the depolarizing fiber.

With the optical axis of the half-wave plate HWP2 oriented along the diagonal basis, we launch into the depolarizing fiber a polarization state $(|\neswarrow\neswarrow\rangle + |\nwsearrow\nwsearrow\rangle)/\sqrt{2}$ which fully lies in the triplet subspace. This state yields again almost exclusively triplet coincidences depicted in Fig.~\ref{Fig:GoryiDoliny}(c), similarly to the case of parallel polarizations. In contrast, when the half-wave plate HWP2 is rotated by $45^\circ$ thus preparing the singlet state
$(|\neswarrow\nwsearrow\rangle - |\nwsearrow\neswarrow\rangle)/\sqrt{2}$, the situation reverses and we observe a domination of singlet coincidences over triplet coincidences shown in Fig.~\ref{Fig:GoryiDoliny}(d), an effect not attainable with separable states.

We fitted the experimental curves using pairs of Gaussian functions with the same width and the central location.
Assuming that the constant pedestals for sufficiently delayed photons give the effective efficiencies of detecting two-photon events, we can use the visibility parameters of the fitted Gaussians to determine conditional probabilities of detecting a singlet or a triplet coincidence for a given input state using ideal, lossless detectors with multiphoton resolution. Channel capacities per a photon pair evaluated from these data by optimizing the probability distribution over the input ensemble \cite{CoverThomas}
are 0.3~bit for the separable case and 0.82~bit for the entangled case.

In conclusion, we demonstrated experimentally that the application of entangled states allows one to encode more information in the polarization degree of freedom by improving distinguishability of the states emerging from the noisy channel. Polarization is of course only one of the available degrees of freedom for electromagnetic fields that in the most general scenario needs to be considered jointly with other observables such as frequency, photon number, and phase \cite{GaussianChannels}. The enhancement gained by entangling inputs across multiple channel uses is another example of distinctness between the classical and the quantum theories of communication, apart from the nontrivial role played in the latter one by collective measurements on multiple channel outputs \cite{CollectiveMeasurements}.

Finally, let us note certain parallels of our scheme to experimental quantum dense coding \cite{MattWeinPRL96} which also uses maximally entangled Bell states to encode classical information. For quantum dense coding however, one of the particles is initially on the receiver end, whereas in our experiment both particles are transmitted from the sender to the receiver with only a short time separation. Additionally, the incomplete Bell-state analyser based on linear optics, which lowered the capacity of practical quantum dense coding below its theoretical limit, is completely sufficient in our application due to the noise intervening in the channel.  

We have benefited from discussions with A. K. Ekert, J. Mostowski, I. A. Walmsley, K. W\'{o}dkiewicz, and M. \.{Z}ukowski. 
This work has been supported by the Polish Committee for Scientific Research (KBN Grant no. PBZ/KBN/043/P03/2001) and is a part of the general program on quantum engineering of the FAMO National Laboratory in Toru\'{n}, Poland. A.D. thanks The Foundation for Polish Science for the support with the Annual Stipend for Young Scientists.

\end{document}